\documentclass[preprint,12pt]{elsarticle}

\usepackage{xcolor}
\usepackage{amssymb}
\usepackage{multirow}
\usepackage{array}
\newcolumntype{L}[1]{>{\raggedright\let\newline\\\arraybackslash\hspace{0pt}}m{#1}}
\newcolumntype{C}[1]{>{\centering\let\newline\\\arraybackslash\hspace{0pt}}m{#1}}
\newcolumntype{R}[1]{>{\raggedleft\let\newline\\\arraybackslash\hspace{0pt}}m{#1}}
\usepackage{amsmath,amssymb,amsfonts}

\usepackage[swedish,english]{babel}

\begin{document}

\begin{frontmatter}

\title{Novel entropy difference-based EEG channel selection technique for automated detection of ADHD}

\author[inst1]{Shishir Maheshwari \corref{cor1}\fnref{fn1}}
\ead{shishir.maheshwari@thapar.edu, shishirweb@gmail.com}
\author[inst2]{Kandala N V P S Rajesh }
\ead{rajesh.k@vitap.ac.in}
\author[inst3]{Vivek Kanhangad}
\ead{kvivek@iiti.ac.in}
\author[inst4]{U Rajendra Acharya}
\ead{Rajendra.Acharya@usq.edu.au}
\author[inst5]{T Sunil Kumar \fnref{fn1}} 
\ead {sunilkumar.telagam.setti@hig.se}
\affiliation[inst1]{organization={Department of Electronics and Communication Engineering, Thapar Institute of Engineering \& Technology},
            city={Patiala},
            postcode={147004}, 
            state={Punjab},
            country={India}}
\affiliation[inst2]{organization={School of Electronics Engineering, VIT-AP University},
            city={Vijayawada},
            postcode={522237}, 
            state={Andhra Pradesh},
            country={India}}
\affiliation[inst3]{organization={Department of Electrical Engineering, IIT Indore},
            city={Indore},
            postcode={453 552}, 
            state={Madhya Pradesh},
            country={India}}
\affiliation[inst4]{organization={School of Mathematics, Physics, and Computing, University of Southern Queensland },
            city={Springfield},
            country={Australia}  } 
 \affiliation[inst5]{organization={Department of Electrical Engineering, Mathematics and Science, University of Gävle},
            city={Gävle},            
            country={Sweden}, 
             }           
\cortext[cor1]{Corresponding author}  
\fntext[fn1]{Shishir Maheshwari and T Sunil Kumar have contributed equally and Share first-authorship}

\begin{abstract}
    Attention deficit hyperactivity disorder (ADHD) is one of the common neurodevelopmental disorders in children. This paper presents an automated approach for ADHD detection using the proposed entropy difference (EnD)-based encephalogram (EEG) channel selection approach. In the proposed approach, we selected the most significant EEG channels for the accurate identification of ADHD using an EnD-based channel selection approach. Secondly, a set of features is extracted from the selected channels and fed to a classifier. To verify the effectiveness of the channels selected, we explored three sets of features and classifiers. More specifically, we explored discrete wavelet transform (DWT), empirical mode decomposition (EMD) and symmetrically-weighted local binary pattern (SLBP)-based features. To perform automated classification, we have used k-nearest neighbor (k-NN), Ensemble classifier, and support vectors machine (SVM) classifiers. Our proposed approach yielded the highest accuracy of 99.29\% using the public database. In addition, the proposed EnD-based channel selection has consistently provided better classification accuracies than the entropy-based channel selection approach. Also, the developed method has outperformed the existing approaches in automated ADHD detection.

\end{abstract}

\begin{keyword}
    Attention deficit hyperactivity disorder (ADHD), entropy difference (EnD)-based EEG channel selection, empirical mode decomposition (EMD), symmetrically-weighted local binary pattern (SLBP).
\end{keyword}

\end{frontmatter}

\section{Introduction}
   Attention deficit hyperactivity disorder (ADHD) is one of the neurodevelopmental disorders in children. ADHD is characterized by less attention, impulsive nature, and hyperactivity \cite{1}. Generally, it originates in childhood and lasts up to adulthood \cite{2}. The most common symptoms are inattentiveness, repeatedly making the same mistakes, facing difficulty performing tasks and activities, talking too much, or combining inattentive symptoms and hyperactivity symptoms \cite{1}. According to the diagnostic and statistical manual of mental disorders (DSM), it is reported that ADHD affects approximately 3\%-5\% of school children, and comparatively, boys have more prevalence than girls \cite{3}. In \cite{joseph2019prevalence}, the authors conducted a meta-analysis to estimate the pooled prevalence of ADHD in India. The meta-analysis showed that the overall prevalence of ADHD in India was 7.1\% (95\% confidence interval: 5.5\%-8.8\%). The prevalence was higher in boys (8.8\%) than in girls (5.1\%). Extensive studies state that there is interdependence between ADHD and substance use and abuse, especially in adolescence. Also, lack of attention and impulsivity creates an overburden to the parents, family, and society \cite{4}. Therefore, the early recognition of ADHD may help provide proper therapy and, thereby, better living.     
    
   Currently, clinicians follow the guidelines of the DSM-fourth edition (DSM-IV) and fifth edition (DSM-V) for ADHD diagnosis \cite{5,loh2022automated}. This diagnosis process depends on the analysis of the subject's symptoms, discussion with parents, and based on a questionnaire. However, this process delivers some inconsistent measures because of its subjective assessment~\cite{faraone2006age}. Therefore, there is a requirement for a quantitative analysis that can differentiate ADHD from non-ADHD subjects. As ADHD is related to inhibitory control's cognitive or learning impairment, a neurophysiological test can give good results \cite{6}. Electroencephalogram (EEG) is a relatively inexpensive, non-invasive, and efficient method that can be used to study the cognitive changes in the brain \cite{7}. 
    
    In the past few decades, several researchers have focused on ADHD detection. As a result, several approaches have been proposed in the literature \cite{8,9,10,11,12,13}. Lubar et al. explored EEG for ADHD detection \cite{8} in 1991. The studies show that the theta frequency band power to beta frequency band power is higher in ADHD than in healthy controls. In \cite{5}, the authors have computed auto-regressive coefficients-based features and are fed to k-nearest neighbor (k-NN) classifier for ADHD detection. They have selected the most discriminated EEG channels based on the accuracy and area under the curve (AUC) values. The proposed method achieved 90\% accuracy and 0.98 AUC values. However, only eight subjects of EEG data are used for training and testing, which needs to be increased for more generalization. In \cite{11}, nonlinear features, namely, Lyapunov exponent, Higuchi fractal dimension, Katz fractal dimension, and Sevcik fractal dimension, are computed from EEG channels as features, and a multi-layer perception (MLP) is used for classification. This work also proposed an EEG channel selection based on the region of electrode lobes. It stated that for given features and classifier, the frontal region of EEG yields better performance of 96.7\% accuracy. The authors' approach to randomly selecting training and testing groups without proper validation methods may not adequately evaluate the model's performance. This may lead to data leakage also ~\cite{hastie2013introduction,kuhn2013applied,geron2022hands,shalev2014understanding}.
    
    The approach proposed in \cite{12} is also based on the nonlinear features: fractal dimension (FD), approximate entropy, and Lyapunov exponent and MLP classifier. Here, a minimum Redundancy Maximum Relevance (mRMR) scheme is used for obtaining the most significant features from the whole feature set, and the approach obtained 93.65\% accuracy. The major limitation of this work is that the subjects used in this study have not undergone any cognitive assessment to say about any cognitive differences between the two classes of subjects. In recent work, \cite{13}, two different decomposition methods, namely wavelet decomposition and empirical mode decomposition, are used as an initial feature extraction step. Later, autoregressive model coefficients, relative wavelet energy, and nonlinear features are extracted, followed by a feature selection method. The sequential forward selection method is employed to reduce the features. Finally, the features are subjected to a k-NN classifier and obtained 97.8\% accuracy. This scheme is evaluated on 123 subjects of ADHD and control groups of children. In~\cite{SHARMA2023119219}, three multivariate decomposition techniques: multivariate empirical mode decomposition (MEMD), multivariate empirical wavelet transform (MEWT), and multivariate variational mode decomposition (MVMD), are used for ADHD detection. After individually decomposing the EEG signals into subbands using the three methods mentioned above, they computed their instantaneous amplitudes and frequencies.
    A total of 150 features were extracted, and later using feature selection methods, this feature set was reduced to 15. Using the leave-one-subject-out validation scheme, the authors reported 92 \% accuracy when features extracted from MVMD were subjected to the artificial neural network (ANN). The authors mentioned that the EEG channel selection-based scheme could be a potential feature extraction scheme they plan in the future. In~\cite{khare2022vhers}, the authors proposed a new algorithm called  variational mode decomposition and Hilbert transform-based EEG rhythm separation (VHERS) that combines variational mode decomposition (VMD) and Hilbert transform (HT) for separating the EEG signals into different rhythms. The VMD algorithm decomposes the EEG signals into a set of intrinsic mode functions (IMFs), which are then transformed using the HT to extract each rhythm's amplitude and phase information. The authors use this information to extract features that capture the differences between ADHD and healthy control subjects. The authors evaluated their approach using a dataset of EEG signals collected from 26 ADHD patients and 26 healthy control subjects. They achieved an accuracy of 94.23\% in distinguishing between the two groups using a support vector machine (SVM) classifier. A novel eight-pointed star pattern learning network (EPSPatNet86)-based ADHD detection is proposed in~\cite{r2}. The authors combined convolutional and recurrent neural networks to analyze the EEG signals and extract relevant features. The authors evaluated the performance of the EPSPatNet86 for ADHD detection. They achieved an accuracy of 93.17\%, in detecting ADHD. A novel ternary motif pattern (TMP)-based ADHD detection model using EEG signals was proposed in~\cite{r1}. The authors used TMP to extract features from the EEG signals and then fed them into a machine-learning algorithm for classification. Directed phase transfer entropy (dPTE)-based effective connectivity matrices (ECM) are computed from the EEG signals for identifying children with ADHD ~\cite{r3}. An effective connectivity vector (ECV) is created as a feature vector from ECM computed from different frequency bands of the EEG signals. Later, this feature vector is fed to MLP for classification. The study suggests that analyzing dPTE values from EEG signals may provide a reliable approach to identifying children with ADHD. The authors in ~\cite{r4} decomposed the preprocessed EEG signals into four frequency bands ( theta, alpha, beta, and gamma). Later, these bands are converted to an RGB image. Finally, the images are processed using a convolutional neural network (CNN) to detect ADHD from EEG. The authors in~\cite{r5} proposed an approach for diagnosing ADHD using nonlinear EEG signal analysis. They computed nonlinear features, including fractal dimension, Hurst exponent, and correlation dimension. The study hypothesizes that nonlinear EEG signal analysis can provide more sensitive and specific measures for diagnosing ADHD than traditional linear measures. The authors in~\cite{r6} proposed a nonlinear causal relationship estimation by an artificial neural network (nCREANN) method to distinguish between ADHD and healthy children. The nCREANN method estimates the linear and nonlinear patterns of the EEG signal for discriminating the ADHD and healthy subjects. 
    
    Channel selection is a process of selecting the most relevant EEG channels that are suitable for an application. Reducing the channels required for automated diagnosis not only reduces the complexity but also helps improve the approach's performance. Also, this aids in developing portable EEG acquisition systems that may be more convenient for patients~\cite{lotte2018review,schirrmeister2017deep}. Despite the advantages, most of the existing approaches in ADHD detection have not explored channel selection. The approach in \cite{5} has selected the effective and discriminative channels based on the accuracy obtained in ADHD detection. The approach in \cite{11} has investigated the performance channels from different lobe regions for ADHD detection. The authors in~\cite{khare2023explainable} developed a model that combines feature selection, decision tree, and logical rule extraction techniques to extract relevant features from the EEG signals and generate an interpretable model for ADHD detection. The authors also analyzed the interpretable model generated by their approach to identify the most relevant EEG features for ADHD detection. They found that the interpretability and explainability of the frontal region are highest compared with the other brain regions. In spite of the success of the above-mentioned approaches, they have the following limitations.
    \begin{itemize}
        \item Recent studies on ADHD detection using EEG are developed based on the different deep learning models. Although the results are promising, these algorithms add complexity to the models for real-time implementation~\cite{r1,r4}.
        \item Despite many advantages of EEG channel selection including reduced computational complexity, and reduced setup time~\cite{alotaiby2015review}, very few approaches~\cite{5,11,khare2023explainable} have explored channel selection in ADHD detection. These approaches, select EEG channels empirically based on performance metrics, which might not necessarily identify the most informative channels. A systematic approach for channel selection is essential for enhancing both accuracy and interpretability.
        
    \end{itemize}
    
    The main contributions of this work are as follows:
    
    \begin{itemize}
        \item Introduced a channel selection approach based on the entropy difference (EnD) value.
        \item Developed an automated approach for detecting ADHD using the significant channels selected using the EnD-based channel selection approach.
        \item Demonstrated the effectiveness of the proposed ADHD framework with three different feature sets using three machine learning algorithms. Also, the proposed approach has outperformed the existing approaches in ADHD detection.
    \end{itemize}

    The rest of the paper is organized as follows: Section~\ref{sec.2} explains the proposed EnD-based ADHD detection approach. Section~\ref{sec.3} presents results obtained and finally, conclusions are drawn in Section~\ref{sec.4}.

\section{Methodology}
\label{sec.2}
   This section briefly describes the proposed EnD-channel selection-based ADHD approach.
   
    \begin{figure*}[!h]
        \centering
        \centerline{\includegraphics[scale=.65]{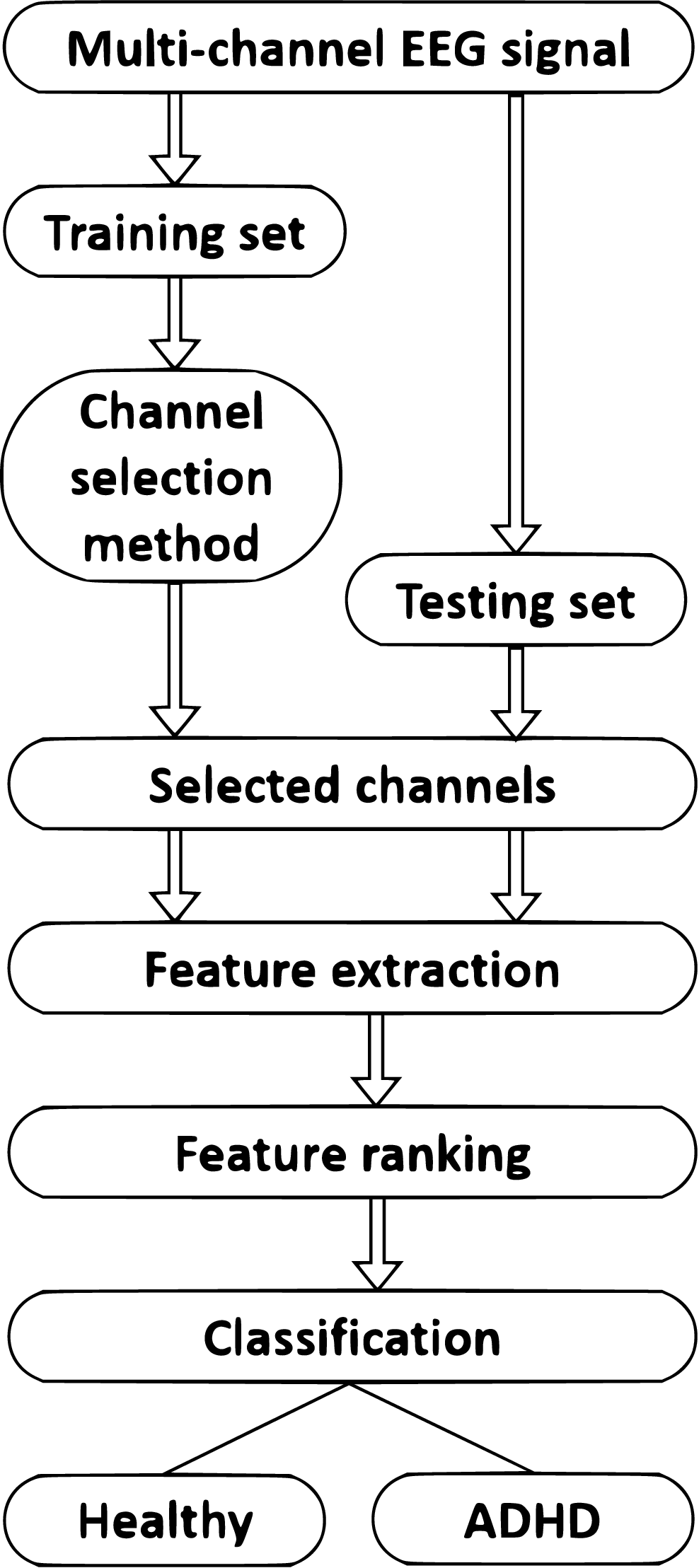}}
        \caption{Block diagram of the proposed methodology for the automated detection of ADHD detection from EEG signals. }
        \label{fig1}
    \end{figure*}

    \begin{figure*}[!h]
        \centering
        \centerline{\includegraphics[width=210mm,height=170mm]{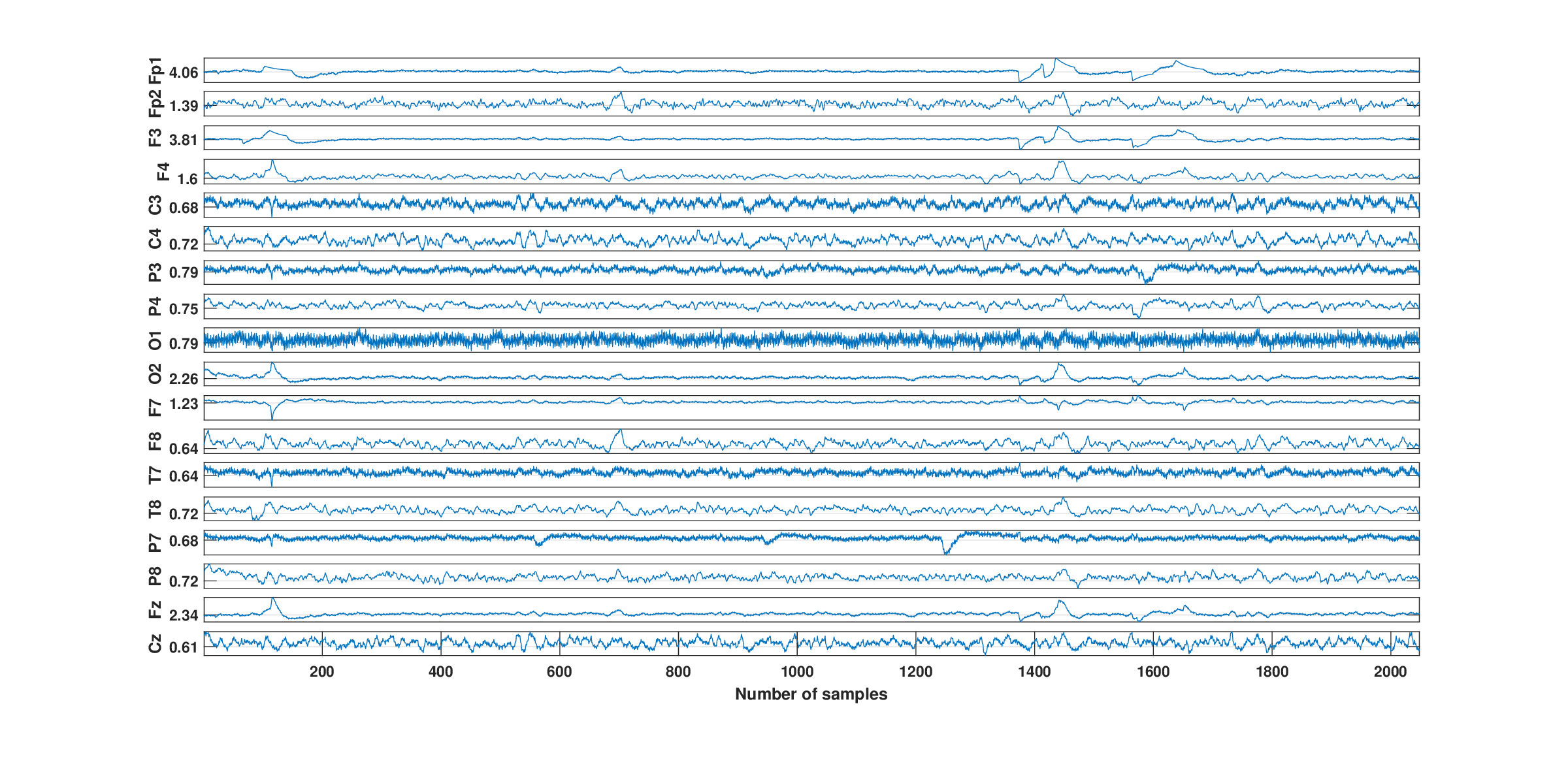}}
        \caption{Sample plot of multi-channel EEG segment. }
        \label{figsig}
    \end{figure*}

   The block diagram of the proposed ADHD detection approach using EnD-based channel selection is shown in Fig.~\ref{fig1}. Fig.~\ref{figsig} shows the sample multi-channel EEG signal. Firstly, the multi-channel EEG data is divided into train and test sets. Then, the significant channels are selected from the train data based on EnD value, and three sets of features are extracted from each channel. Further, the same set of selected channels is employed in the test set. The features corresponding to individual channels are concatenated to get a final feature vector. After that, the most discriminative features are selected from the extracted feature vector. Finally, to test the performance of the proposed approach, the selected features are fed to three different classifiers to detect ADHD automated.
    
    Furthermore, we have also used entropy-based channel selection to test the effectiveness of our proposed feature selection approach. This method also employed a similar approach as proposed for EnD-based channel selection. The proposed approach is further detailed below.

    
    
    \subsection{Entropy difference (EnD) based method for channel selection}
         Entropy-based channel selection: Before introducing EnD-based channel selection, we briefly review the entropy-based channel selection. As the name suggests, here the channel selection is based on entropy value. Entropy is a measure of randomness or uncertainty of an EEG channel~\cite{shannon2001mathematical}. The entropy of channel $C$ is given as: 

        \begin{equation}
            H(C)= - \sum_{i=1}^{n} p(x_i) log_2 p(x_i)
        \end{equation}

        $p(x_i)$ is the probability mass function of the EEG channel with $n$ samples. After computing entropies of all the EEG channels, the first $N$ channels with the highest entropy are selected for ADHD detection.

        \begin{table}[h!]
                \centering
        \caption{Ranking obtained using the entropy-based channel selection approach.}
\begin{tabular}{ c c | c c | c c | c c } 
                     \hline
                     S. No. & Channel & S. No. & Channel & S. No. & Channel & S. No. & Channel \\
                            & name   &        & name   &        & name   &        & name\\
                     \hline
                     1 & Cz & 6 & O2 & 11 & Pz & 16 & F7 \\
                     2 & F4 & 7 & P7 & 12 & T8 & 17 & Fp2 \\
                     3 & P8 & 8 & F3 & 13 & F8 & 18 & Fz \\
                     4 & P4 & 9 & O1 & 14 & P3 & 19 & Fp1 \\
                     5 & C4 & 10 & C3 & 15 & T7 &  &  \\
                     \hline
                \end{tabular}
                \label{t-ch-en}
        \end{table}
        
        EnD-based channel selection: To select the most effective channels for ADHD detection, we develop EnD-based channel selection. The entropy difference of channel $C$ is computed as follows:

        \begin{equation}
            EnD(C) = abs(H_{ADHD}(C) - H_{HC}(C))
        \end{equation}

        After computing EnD, we selected the top $N$ channels based on the maximum entropy difference. Here, $H_{ADHD}(C)$ represents the entropy of EEG channel $C$ of ADHD in training data, and $H_{HC}(C)$ represents the entropy of EEG channel $C$ of healthy control in training data. The significance of the EnD is that it represents the change in information due to ADHD. More specifically, high EnD for a channel represents a significant change in information due to ADHD, and low EnD represents less change in information. 

        \begin{table}[h!]
                \centering
                \caption{Ranking obtained using the EnD-based channel selection approach.}
                \begin{tabular}{ c c | c c | c c | c c } 
                     \hline
                     S. No. & Channel & S. No. & Channel & S. No. & Channel & S. No. & Channel \\
                            & name   &        & name   &        & name   &        & name\\
                     \hline
                     1 & Fz & 6 & P3 & 11 & P4 & 16 & Fp2 \\
                     2 & Pz & 7 & F3 & 12 & T8 & 17 & C4 \\
                     3 & P7 & 8 & Fp1 & 13 & Cz & 18 & F4 \\
                     4 & O1 & 9 & C3 & 14 & O2 & 19 & P8 \\
                     5 & T7 & 10 & F7 & 15 & F8 &  &  \\
                     \hline
                \end{tabular}
                
                \label{t-ch-end}
        \end{table}
\subsection{Feature Extraction}
This section presents brief details about the extracted features from the EEG segments.
    \subsubsection{Discrete wavelet transform-based features}
       The discrete wavelet transform (DWT) is a mathematical technique that provides a time-frequency representation of signals~\cite{mallat1999wavelet}. In the wavelet transform, a signal is decomposed into a series of wavelet coefficients representing different scales and positions. The wavelet coefficients at each scale and position provide information about the frequency content of the signal at a particular time. It allows the wavelet transform to simultaneously capture a signal's time and frequency-domain features.



        The db4 wavelet has become a popular choice for wavelet transforms due to its good balance between temporal and frequency resolution and its ability to handle signals effectively with abrupt changes \cite{wave1},\cite{wave2}. The EEG segments are decomposed into three decomposition levels using the db4 mother wavelet in this proposed approach. Further from each sub-component, four statistical features such as Shannon entropy, mean, variance, and energy are extracted.
        
    
    \subsubsection{EMD-based features}
        EMD is a time-frequency technique that decomposes a non-stationary signal into a finite number of modes  known as IMFs~\cite{huang1998empirical}. Each IMF consists of a narrow band frequency and is extracted without any prior knowledge of the underlying frequencies. Unlike wavelets, EMD decomposes signals adaptively based on the frequency component of the signal. It also does not employ any basis function as used by wavelets. A signal $S(t)$ can be decomposed into IMFs as follows:

        \begin{equation}
            S(t) = \sum_{n=1}^{N} \mathrm(IMF)_{n}(t) + Res_N(t)
        \end{equation}

        where $S(t)$ is the multi-component signal. $IMF_{n}(t)$ is the $n^{th}$ IMF and $Res_N(t)$ represents the residue corresponding to $N$ intrinsic modes. Each IMF satisfies two conditions: (1) the number of extrema and zero-crossings must be equal or differ by at most one, and (2) the mean value of the envelope defined by the local maxima and minima must be zero. More mathematical details of EMD can be found in~\cite{rilling2003empirical}.



        In this work, the EEG segments are decomposed into 3 IMFs using EMD. As mentioned in the above sub-section, the same set of statistical features is extracted from these IMFs.
    
    \subsubsection{SLBP-based features}
        SLBP-based features are time-domain features. It is a local descriptor, which is found to be effective in the classification of EEG signals~\cite{rajesh2021schizophrenia}. In SLBP computation, a binary string is generated for every sample in a signal. The binary string is generated by performing comparisons in a sample's right and left neighborhoods. Thus, a binary string is converted into a decimal number using a symmetrical weighting scheme. By computing SLBP-based features for all the samples in a signal, we can obtain a feature vector that captures the local variations of the signal. SLBP is mathematically computed as follows:
\begin{equation}
\mathrm{SLBP}_{LHS}(x[n])=\sum_{m=0}^{L-1} f(x[n+m-L]-x[n]) 2^{L-1-m}
\end{equation}
\begin{equation}
\mathrm{SLBP}_{RHS}(x[n])=\sum_{m=0}^{L-1} f(x[n+m+1]-x[n]) 2^{L}
\end{equation}
\begin{equation}
\mathrm{SLBP} (x[n])= \mathrm{SLBP}_{LHS} (x[n])+ \mathrm{SLBP}_{RHS}(x[n])
\end{equation}
where $x[n]$ represents the current sample of the signal, $L$ is the neighborhood size considered on either side of the current sample, and threshold function $f$ is defined as 
\begin{equation}
f(x)=\left\{\begin{aligned}
0, && x<0 \\
1, && x \geq 0
\end{aligned}\right.
\end{equation}
Sample SLBP computation can be found in~\cite{sunil2017automated}.
From each EEG segment, the SLBP provides 31 features. 

    \subsection{Feature selection using Chi-square test}
        The performance of the computer-aided system significantly depends on the features extracted from the data. Each feature does not contribute equally, and features with low inter-class discriminating capability hamper the system's performance. The feature selection method provides each feature's discriminating value, which can be used to select valuable features. In this work, we employed a Chi-square test for feature selection~\cite{zhai2018chi}. The Chi-square test evaluates the relationship between each feature and the response variable (i.e., the class label). The test measures the dependence of the feature and the response by comparing the observed frequency distribution of the feature for each class to the expected frequency distribution under the assumption of independence. If the Chi-square statistic exceeds the critical value, the feature is considered to have a significant relationship with the response and is selected as a discriminatory feature~\cite{zhai2018chi}.

  \subsection{Classifiers}
    This subsection details the machine learning algorithms used for automated detection of ADHD using EEG signals. 
    
    \subsubsection{Support Vector Machine (SVM)}
        SVM is a supervised classifier proposed by Vapnik~\cite{vapnik1998statistical}. SVMs are naturally binary classifiers, and they construct a hyperplane in the higher dimensional feature space to separate the boundaries between feature vectors from both classes. Finding the best hyperplane that gives the highest margin distance between the nearest feature vectors of the two class labels is an optimization problem. The main advantage of this algorithm is that it can also be extended to multiclass classification by following one-against-all or one-against-one techniques. By using a kernel method, nonlinear data can also be separated in a higher dimensional space. More details can be found in~\cite{ duda2006pattern}.

        \subsubsection{k-nearest neighbor (k-NN)}
        k-NN is another supervised learning algorithm like SVM. Unlike SVM, kNN is instance-based learning. It means k-NN uses whole data at a time for training instead of learning some parameters like in SVM. Hence, k-NN is also known as non-parametric learning. k-NN stores the entire dataset in the memory and tests a new sample (feature vector) with k nearest (closest in the distance) training feature vectors in the memory and allocates the class label using a majority voting scheme. The often-used distance metrics are Euclidean distance and cosine similarity. More details can be found in~\cite{burkov2019hundred}.

        \subsubsection{Ensemble Learning}
        Ensemble learning models are suitable for nonlinear data processes ~\cite{ burkov2019hundred}. The ensemble learning method trains various low-accuracy models and clubs their predictions to get the final prediction result. It gives better generalization and accurate results and alleviates the problem of finding a unique, accurate model for solving the given problem. Moreover, learning a weak model reduces complexity and increases the speed of the process. More details of ensemble learning models can be found in~\cite{burkov2019hundred}.

\section{Results}
\label{sec.3}
    This section presents the details about the dataset employed, and the experimental results.

    \subsection{Dataset}
        The dataset employed to assess the performance of the proposed approach is obtained from the IEEE dataport~\cite{nasrabadi2020eeg,samavati2012automatic}. The study comprised 121 children (boys and girls, ages 7-12), with 61 children diagnosed with ADHD and 60 healthy controls. The sample EEG signal is shown in Fig.~\ref{figsig}. The ADHD diagnosis for the children was confirmed by an experienced psychiatrist using the DSM-IV criteria. They had been taking Ritalin (medicine for treating ADHD) for a maximum of 6 months. On the other hand, none of the children in the control group had any history of psychiatric disorders, epilepsy, or high-risk behaviors. EEG recording was conducted using a 19-channel system, with electrode placements following the 10-20 standard. The recording included 19 channels, specifically Fz, Cz, Pz, C3, T3, C4, T4, Fp1, Fp2, F3, F4, F7, F8, P3, P4, T5, T6, O1, and O2, at a sampling frequency of 128 Hz. The A1 and A2 electrodes located on the earlobes served as the reference electrodes for the recording wth a sampling frequency of 128 Hz. 

        The EEG recording protocol utilized for data recording is visual attention tasks. In the task, children were presented with pictures featuring cartoon characters and asked to count the number of characters in each image. The number of characters varied randomly between 5 and 16, and the pictures were sized appropriately to allow for easy counting. To maintain a continuous stimulus during EEG recording, each image was displayed without interruption immediately after the child's response. Consequently, the duration of the EEG recording for this cognitive visual task depended on the child's response speed.
    
 \subsection{Experimental Results}
In this section, we presented the experimental conditions and results of our proposed method in Figure~\ref{fig1}. To demonstrate the effectiveness of the EnD-based channel selection approach, we utilized three distinct feature extraction methods: EMD, DWT, and SLBP-based feature extraction approaches.

To identify ADHD from EEG signals, we employed three different supervised classifiers: SVM with RBF kernel, k-NN, and Ensemble Learning (ENS).

To evaluate the performance of our approach, we utilized three validation strategies and are given below:

\begin{enumerate}
    \item 70:30 split ratio: The dataset was split into a 70\% training set and a 30\% testing set. The classification model was trained on the training set and evaluated on the testing set.
    \item 70:30 random split evaluation: The 70:30 split evaluation was repeated ten times, each time with a different random split. The results from these ten evaluations were averaged to obtain a more robust and representative performance measures.
    \item Ten-fold cross-validation: The dataset was divided into ten folds, with each fold serving as a testing set once while the remaining nine folds are used for training. This process was repeated ten times, ensuring that each fold acted as the testing set once. The performance metrics obtained from these ten iterations were combined to assess the classification model's overall effectiveness.
\end{enumerate}

In this work, we have used accuracy, sensitivity and specificity parameters to evaluate the performance of the proposed ADHD approach~\cite{r4}.\\
    
\subsubsection{70:30 split ratio} 
In the 70:30 split evaluation setup, firstly, the dataset is divided into training (the first 70\% of the data) and testing (the last 30\% of the data) subsets. After that, we divided each channel into segments of  2048 (16 seconds) samples. As the EEG signals used in this study are collected from multiple channels, capturing information from different lobes of the brain. Each channel provides distinct brain information. Consequently, we conducted a channel-based analysis for ADHD diagnosis. To accomplish this, the first step in our method is to assign ranks to all 19 channels of each EEG signal. Two approaches were employed: entropy-based channel ranking and entropy difference-based channel ranking techniques. The ranks obtained from these two methods, ranging from high (1) to low (19), are presented in Tables~\ref{t-ch-en} and \ref{t-ch-end}.

\begin{figure*}[!h]
            \centering
            \centerline{\includegraphics[scale=.55]{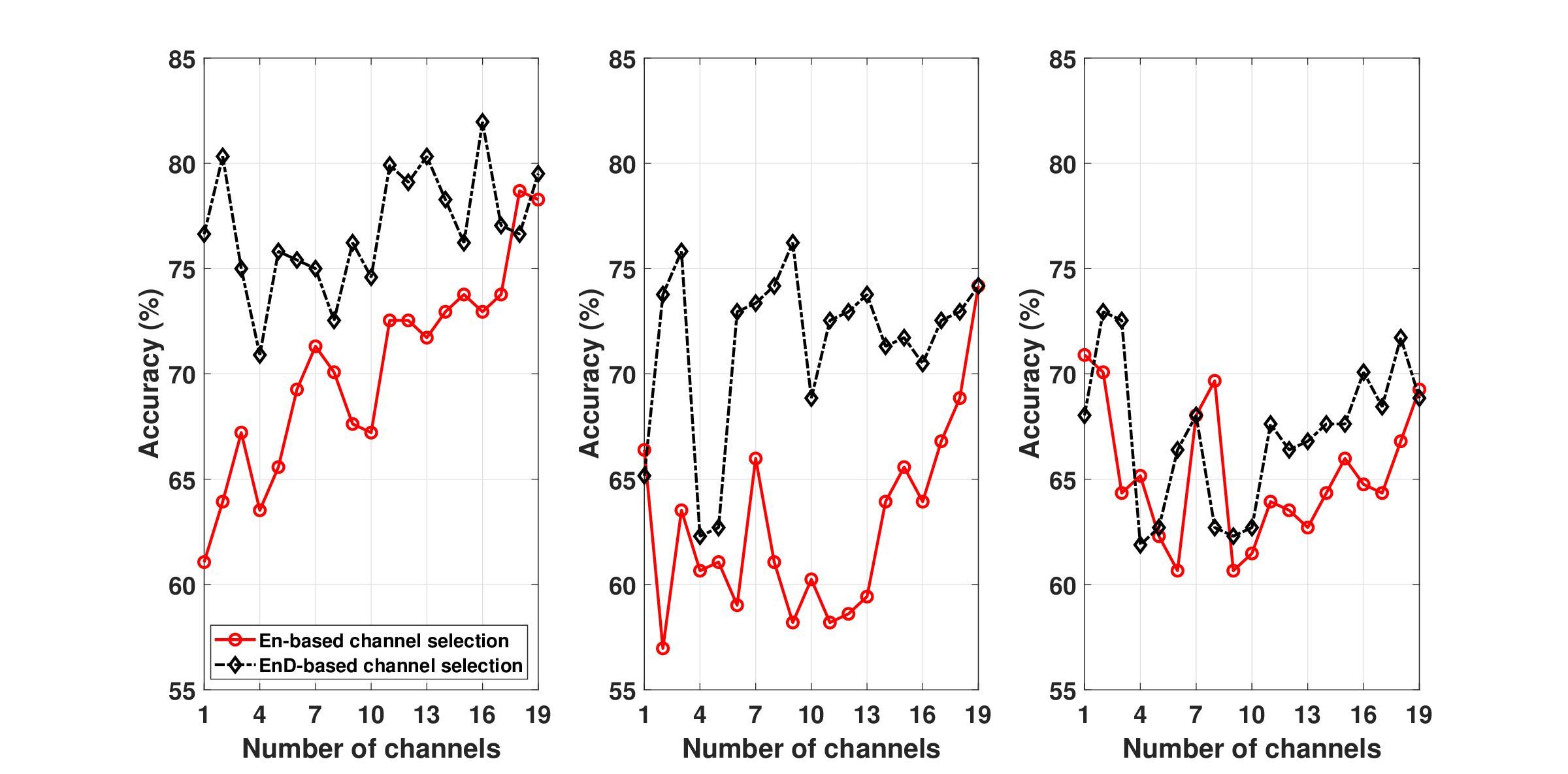}}
            \caption{Comparison of ENS (left), k-NN (middle), and SVM (right) classifier accuracy obtained from EMD-based features for EN- and EnD-based channel selection method (before feature selection).}
            \label{fig_allch_emd}
            \end{figure*}

\begin{figure*}[!h]
            \centering
            \centerline{\includegraphics[scale=.55]{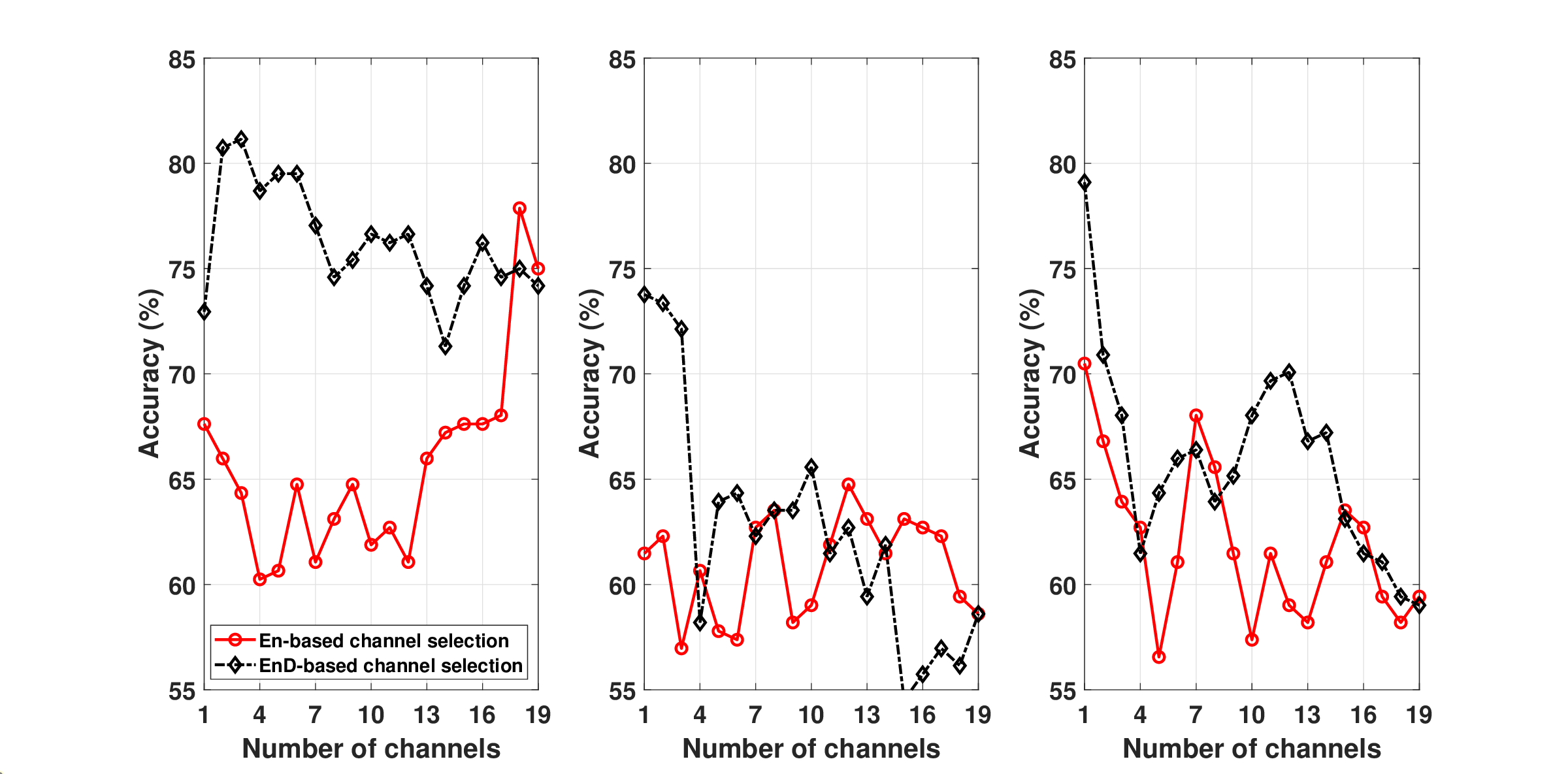}}
            \caption{Comparison of ENS (left), k-NN (middle), and SVM (right) classifier accuracy obtained from DWT-based features for En- and EnD-based channel selection method (before feature selection).}
            \label{fig_allch_wav}
            \end{figure*}

\begin{figure*}[!h]
            \centering
            \centerline{\includegraphics[scale=.55]{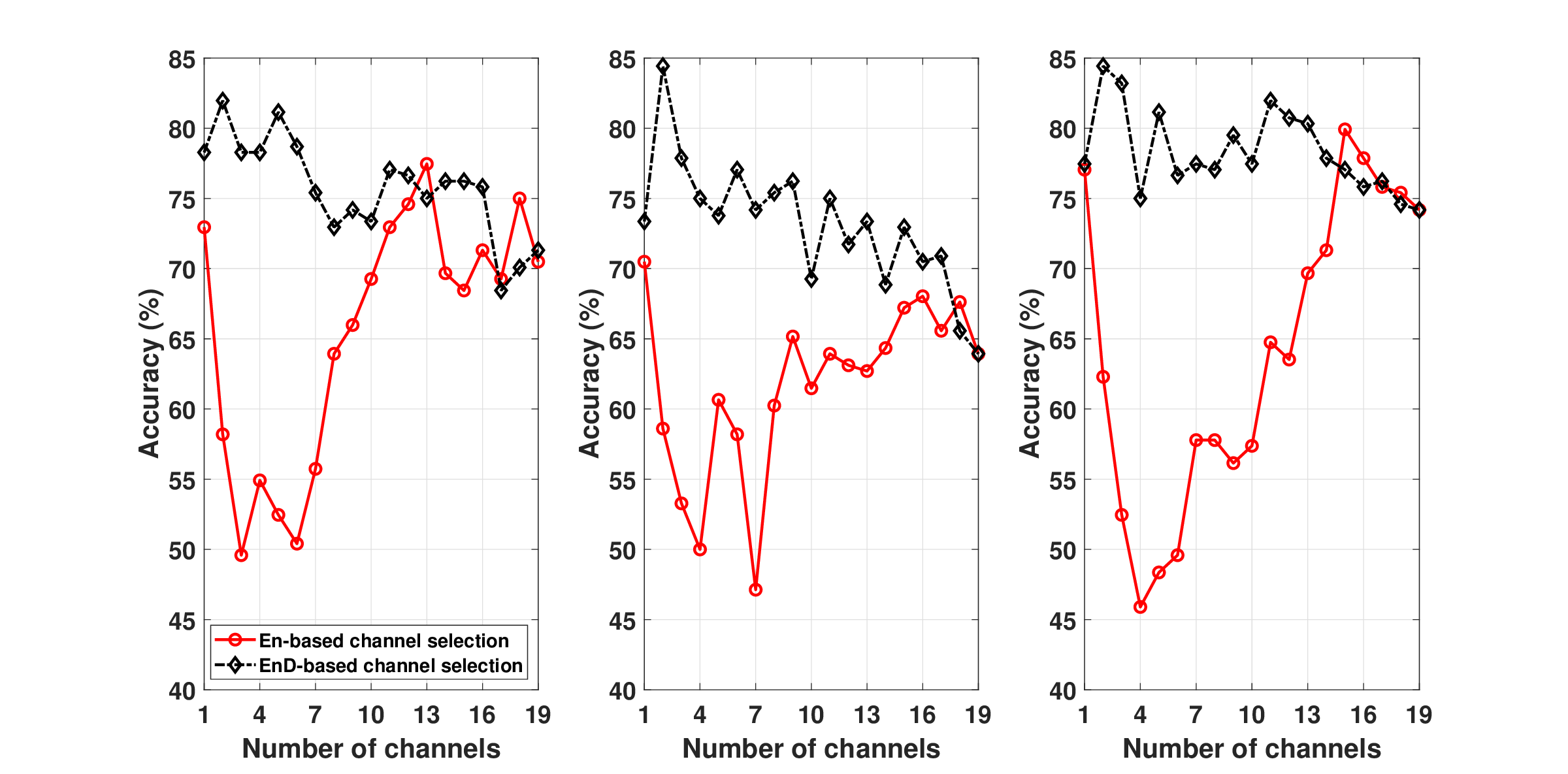}}
            \caption{Comparison of ENS (left), k-NN (middle), and SVM (right) classifier accuracy obtained from SLBP-based features for En- and EnD-based channel selection method (before feature selection).}
            \label{fig_allch_slbp}
            \end{figure*}

After obtaining the channel ranking to compare the EnD and entropy-based channel selection approaches, we plotted the classification accuracy against the number of channels used in ADHD detection. The results obtained are depicted in Figures ~\ref{fig_allch_emd}, \ref{fig_allch_wav}, and \ref{fig_allch_slbp}. Our primary focus was to analyze the performance of the EnD and entropy-based channel selection approaches, so no additional feature selection was applied to these results.
From the figures, it is evident that the channel ranking obtained using the EnD-based method consistently outperformed the entropy-based method across all employed feature extraction and classification methods. Notably, we observed that after the first three best-ranked channels, there was no significant improvement in classification accuracy in the case of the EnD-based channel selection scheme. Furthermore, increasing the number of channels would result in longer feature vectors, potentially leading to increased complexity.

To facilitate further analysis, we considered the feature vectors derived from the first three best-ranked channels. The lengths of the features computed from the selected channels using the various methods mentioned above are presented in Tables ~\ref{t-para-emd}, \ref{t-para-wav}, and \ref{t-para-slbp}. The decomposition level and number of intrinsic mode functions (IMFs) for wavelet transform and EMD, respectively is fixed at 3.

Therefore, for further analysis, we considered the feature vectors from the first three best-ranked channels. To further reduce feature vector length, features extracted from the selected channels are given to the Chi-square test-based feature selection approach. These features are then fed into various supervised classifiers, namely SVM (RBF kernel), k-NN, and ENS, to discriminate between ADHD and healthy subjects. The performance measures for this experimental setup are presented in Tables~\ref{t-acc-emd}, \ref{t-acc-wav}, and \ref{t-acc-slbp}. Tables ~\ref{t-acc-emd}, ~\ref{t-acc-wav} and ~\ref{t-acc-slbp} represent the feature extraction using the EMD, wavelets, and SLBP-based methods, respectively. 

Overall, the comparison of EnD and entropy-based channel selection approaches showed that EnD consistently outperformed entropy channel selection in ADHD detection across different feature extraction and classification methods. Additionally, limiting the feature vectors to the top three ranked channels offered a good balance between accuracy and complexity, leading to more efficient and effective ADHD detection. 
       
            \begin{table}[h!]
                \centering
                \caption{Feature vector length used for EMD-based method }
                \begin{tabular}{ c | c } 
                     \hline
                     Number of channels (NCh) & Feature length \\
                     \hline \hline
                     1 & 12 \\ 
                     2 & 24 \\ 
                     3 & 36 \\ \hline
                \end{tabular}
                
                \label{t-para-emd}
            \end{table}
            
            \begin{table}[h!]
                \centering
                \caption{Feature vector length used for Wavelet-based method.}
                \begin{tabular}{ c |  c } 
                     \hline
                     NCh & Feature length \\
                     \hline \hline
                     1 & 16 \\ 
                     2 & 32 \\ 
                     3 & 48 \\ \hline
                \end{tabular}
                
                \label{t-para-wav}
            \end{table}
            
            \begin{table}[h!]
                \centering
                \caption{Feature vector length used for SLBP-based method.}
                \begin{tabular}{ c | c } 
                     \hline
                     Nch & Feature length \\
                     \hline \hline
                     1 & 31 \\ 
                     2 & 62 \\
                     3 & 93 \\ \hline
                \end{tabular}
                \label{t-para-slbp}
            \end{table}

             \begin{table}[h!]
                \centering
                \caption{Results obtained with EMD-based features using 70:30 split ratio strategy.}
                \resizebox{1.15\textwidth}{!}{
                \begin{tabular}{  c | c | C{2.5cm} | C{2.5cm} | C{2.50cm} | C{2.50cm} } 
                     \hline
                     \multirow{2}{*}{Classifier} & \multirow{2}{*}{NCh} & \multicolumn{2}{c|}{Acc in \%} & \multicolumn{2}{c}{Sn, Sp in \%} \\ \cline{3-6}
                     & & En & EnD & En & EnD \\ \hline \hline
                     
                     \multirow{3}{*}{ENS} & 1 & 65.98 & 80.73 & 61.05, 69.12 & 78.94, 81.87 \\
                     & 2 & 72.95 & 83.61 & 58.94, 81.87 & 85.26, 82.55 \\
                     & 3 & 72.95 & 83.61 & 58.94, 81.87 & 85.26, 82.55 \\ \hline
                     
                     \multirow{3}{*}{k-NN} & 1 & 66.39 & 76.64 & 62.10, 69.12 & 82.10, 73.15 \\
                     & 2 & 65.98 & 76.64 & 64.21, 67.11 & 82.10, 73.15 \\ 
                     & 3 & 67.21 & 76.64 & 54.73, 75.16 & 82.10, 73.15 \\ \hline
                     
                     \multirow{3}{*}{SVM} & 1 & 71.31 & 78.27 & 80.00, 65.77 & 90.52, 70.46 \\
                     & 2 & 72.13 & 81.55 & 80.00, 67.14 & 93.68, 73.82 \\ 
                     & 3 & 71.72 & 81.55 & 73.68, 70.46 & 92.63, 74.49 \\ \hline
                     
                     \multicolumn{6}{l}{\small En: Entropy-based channel selection} \\
                     \multicolumn{6}{l}{\small Acc: Classification accuracy, Sn: Sensitivity, Sp: Specificity} \\
                \end{tabular} }
                
                \label{t-acc-emd}
            \end{table}

            \begin{table}[h!]
                \centering
                \caption{Results obtained with DWT-based features using 70:30 split ratio strategy.}
                \resizebox{1.15\textwidth}{!}{
                \begin{tabular}{  c | c | C{2.50cm} | C{2.50cm} | C{2.50cm} | C{2.50cm} } 
                     \hline
                     \multirow{2}{*}{Classifier} & \multirow{2}{*}{NCh} & \multicolumn{2}{|c|}{Acc in \%} & \multicolumn{2}{c}{Sn, Sp in \%} \\ \cline{3-6}
                     & & En & EnD & En & EnD \\ \hline \hline
                     
                     \multirow{3}{*}{ENS} & 1 & 69.67 & 79.50 & 43.15, 86.57 & 71.57, 84.56 \\
                     & 2 & 67.21 & 88.11 & 47.36,79.86 & 84.21, 90.60 \\
                     & 3 & 64.75 & 87.29 & 47.36, 75.83 & 88.42, 86.57 \\ \hline
                     
                     \multirow{3}{*}{k-NN} & 1 & 65.98 & 75.00 & 55.78, 72.48 & 86.31, 67.78 \\
                     & 2 & 62.29 & 84.83 & 56.84, 65.71 & 81.05, 87.24 \\ 
                     & 3 & 62.29 & 83.60 & 50.52, 69.79 & 81.05, 85.23 \\ \hline
                     
                     \multirow{3}{*}{SVM} & 1 & 77.86 & 85.24 & 83.15, 74.49 & 85.26, 85.23 \\
                     & 2 & 71.72 & 88.52 & 86.31, 62.41 & 97.89, 82.55 \\
                     & 3 & 72.13 & 88.52 & 74.73, 70.46 & 93.68, 85.23 \\ \hline
                \end{tabular} }
                
                \label{t-acc-wav}
            \end{table}
            
            \begin{table}[h!]
                \centering
                \caption{Results obtained with SLBP-based features using 70:30 ratio strategy.}
                \resizebox{1.1\textwidth}{!}{
                \begin{tabular}{  c | c | C{2.50cm} | C{2.50cm} | C{2.50cm} | C{2.50cm} } 
                     \hline
                     \multirow{2}{*}{Classifier} & \multirow{2}{*}{NCh} & \multicolumn{2}{|c|}{Acc in \%} & \multicolumn{2}{c}{Sn, Sp in \%} \\ \cline{3-6}
                     & & En & EnD & En & EnD \\ \hline \hline
                     \multirow{3}{*}{ENS} & 1 & 72.95 & 79.09 & 65.26, 77.85 & 67.36, 86.57 \\
                     & 2 & 66.39 & 85.24 & 57.89, 71.81 & 82.10, 87.24 \\
                     & 3 & 63.52 & 84.83 & 57.89, 67.11 &73.68, 91.94 \\ \hline
                     
                     \multirow{3}{*}{k-NN} & 1 & 70.90 & 73.36 & 70.52, 71.14 & 69.47, 75.83 \\
                     & 2 & 58.60 & 84.42 & 58.94, 58.38 & 83.15, 85.23 \\
                     & 3 & 58.19 & 79.09 & 67.36, 52.34 & 78.94, 79.19 \\ \hline
                     
                     \multirow{3}{*}{SVM} & 1 & 78.27 & 79.91 & 72.63, 81.87 & 72.63, 84.56 \\
                     & 2 & 64.34 & 85.65 & 62.10, 65.77 & 80.00, 89.26 \\
                     & 3 & 59.01 & 85.65 & 46.31, 67.11 & 82.10, 87.91 \\ \hline
                \end{tabular} }
                
                \label{t-acc-slbp}
            \end{table}

\subsubsection{70:30 random split ratio strategy} 
In this evaluation, a different approach is taken compared to the previous 70:30 split ratio strategy. The dataset is randomly divided into training (70\%) and testing (30\%) sets, and this process is repeated ten times. The purpose of this method is to obtain more reliable performance metrics.

The results of these repeated evaluations for EMD-based features, wavelet-based features, and SLBP-based features are presented in Table~\ref{t-10acc}. From this table, it can be observed that the wavelet-based features exhibited the highest accuracy, reaching 95.68\% with ENS classifier. The performance of wavelet-based features with the EnD approach exhibited the potential for effective ADHD classification.

            \begin{table}[h!]
                \centering
                \caption{Average performance metrics obtained with 70:30 random split.}
                \resizebox{1.1\textwidth}{!}{
                \begin{tabular}{  c | c | c | c | c | c | c } 
                     \hline
                     \multirow{2}{*}{Feature} & \multirow{2}{*}{Classifier} & \multirow{2}{*}{NCh} & \multicolumn{2}{c}{Acc in \%} & \multicolumn{2}{c}{Sn, Sp in \%} \\ \cline{4-7}
                     & & & En & EnD & En & EnD \\ \hline \hline
                     
                     \multirow{9}{*}{EMD-based} & \multirow{3}{*}{ENS} & 1 & 75.93 & 79.23 & 68.932, 81.15 & 85.34, 74.30 \\ 
                     & & 2 & 84.42 & 85.38 & 85.58, 83.45 & 89.65, 81.94 \\
                     & & 3 & 85.24 & 85.24 & 82.88, 87.21 & 80.18, 89.47 \\ \cline{2-7}
                     & \multirow{3}{*}{k-NN} & 1 & 67.39 & 75.76 & 59.16, 73.71 & 80.17, 72.22 \\
                     & & 2 & 74.07 & 81.14 & 65.87, 81.25 & 76.57, 84.96 \\ 
                     & & 3 & 75.00 & 83.07 & 73.33, 76.28 & 80.17, 85.41 \\ \cline{2-7}
                     & \multirow{3}{*}{SVM} & 1 & 75.00 & 81.53 & 61.26, 86.46 & 87.93, 76.38 \\
                     & & 2 & 85.55 & 83.92 & 93.65, 78.4722 & 80.35, 86.71 \\
                     & & 3 & 84.81 & 86.30 & 95.23, 75.6944 & 95.23, 78.47 \\ \hline
                     
                     \multirow{9}{*}{DWT-based} & \multirow{3}{*}{ENS} & 1 & 79.50 & 82.08 & 70.27, 87.21 & 83.33, 80.95 \\ 
                     & & 2 & 85.24 & 92.15 & 89.18, 81.95 & 84.82, 97.90 \\
                     & & 3 & 89.75 & 95.68 & 85.58, 93.23 & 91.07, 99.30 \\ \cline{2-7}
                     & \multirow{3}{*}{k-NN} & 1 & 75.41 & 76.92 & 77.19, 73.80 & 80.17, 74.30 \\
                     & & 2 & 77.17 & 86.88 & 77.50, 76.92 & 81.08, 91.72 \\ 
                     & & 3 & 81.11 & 86.88 & 80.95, 81.25 & 81.08, 91.72 \\ \cline{2-7}
                     & \multirow{3}{*}{SVM} & 1 & 79.60 & 85.84 & 73.21, 84.61 & 83.63, 88.07 \\
                     & & 2 & 90.37 & 87.09 & 99.20, 82.63 & 93.80, 81.48 \\
                     & & 3 & 87.40 & 87.05 & 97.61, 78.47 & 97.61, 78.47 \\ \hline
                     
                     \multirow{9}{*}{SLBP-based} & \multirow{3}{*}{ENS} & 1 & 79.66 & 81.27 & 71.84, 85.50 & 72.72, 89.90 \\
                     & & 2 & 81.85 & 85.60 & 78.99, 84.49 & 87.37, 84.41 \\ 
                     & & 3 & 85.08 & 87.65 & 76.47, 93.02 & 79.79, 93.38 \\ \cline{2-7}
                     & \multirow{3}{*}{k-NN} & 1 & 77.17 & 78.73 & 66.99, 84.78 & 71.13, 84.67 \\
                     & & 2 & 82.19 & 88.75 & 86.36, 77.98 & 82.45, 94.44 \\ 
                     & & 3 & 84.78 & 90.39 & 79.16, 89.10 & 91.08, 89.84 \\ \cline{2-7}
                     & \multirow{3}{*}{SVM} & 1 & 84.23 & 83.26 & 77.66, 89.13 & 79.61, 85.71 \\
                     & & 2 & 84.47 & 89.16 & 90.00, 78.89 & 85.96, 92.06 \\ 
                     & & 3 & 88.76 & 91.05 & 90.83, 87.17 & 87.27, 94.11 \\ 
                     \hline
                     
                \end{tabular} }
                
                \label{t-10acc}
            \end{table}

        \subsubsection{10-fold cross-validation}
In addition to the previous evaluation methods, we also conducted $k$-fold cross-validation strategies to assess the effectiveness of our proposed method. In this validation scheme, the dataset was divided into $k$ folds, with one-fold serving as the testing set while the remaining $k-1$ folds were used for training. This process was repeated $k$ times, and the results were aggregated to provide a comprehensive performance measure. In our study, we employed 10-fold cross-validation, dividing the dataset into ten folds.
The performance measures obtained using 10-fold cross-validation strategy are presented in Tables~\ref{t-acc5-emd}- \ref{t-acc5-slbp}, following the same format as the previous evaluation schemes. From these tables, it is evident that the SLBP-based features achieved the highest accuracy, reaching 99.29\% with the k-NN classifier.

            \begin{table}[h!]
                \centering
                \caption{Results obtained with EMD-based features using 10-fold cross-validation strategy.}
                \resizebox{1.1\textwidth}{!}{
                \begin{tabular}{  c | c | C{1.5cm} | C{1.5cm} | C{3.0cm} | C{3.0cm} } 
                     \hline
                     \multirow{2}{*}{Classifier} & \multirow{2}{*}{NCh} & \multicolumn{2}{|c|}{10-fold Acc in \%} & \multicolumn{2}{c}{10-fold Sn, Sp in \%age} \\ \cline{3-6} 
                     & & En & EnD & En & EnD \\ \hline \hline

                     \multirow{3}{*}{ENS} & 1 & 77.92 & 81.32 & 68.96, 84.94 & 75.36, 86.02 \\
                     & 2 & 84.74 & 90.35 & 79.44, 88.87 & 86.06, 93.73 \\
                     & 3 & 90.35 & 93.87 & 86.30, 93.53 & 92.01, 95.33 \\ \hline

                     \multirow{3}{*}{k-NN} & 1 & 73.79 & 73.58 & 70.79, 76.15 & 69.65, 76.71 \\
                     & 2 & 80.13 & 85.33 & 74.91, 84.23 & 79.92, 89.59 \\
                     & 3 & 87.04 & 91.36 & 85.15, 88.53 & 88.58, 93.53 \\ \hline

                     \multirow{3}{*}{SVM} & 1 & 77.21 & 78.70 & 75.33, 78.68 & 75.79, 80.99 \\
                     & 2 & 80.80 & 86.25 & 77.82, 83.14 & 86.52, 86.02 \\
                     & 3 & 83.13 & 88.05 & 83.78, 82.61 & 89.70, 86.75 \\ \hline

                \end{tabular} }
                
                \label{t-acc5-emd}
            \end{table}
            
        \begin{table}[h!]
                \centering
                \caption{Results obtained with Wavelet-based features using 10-fold cross-validation strategy.}
                \resizebox{1.1\textwidth}{!}{
                \begin{tabular}{  c | c | C{1.5cm} | C{1.5cm} | C{3.0cm} | C{3.0cm} } 
                     \hline
                     \multirow{2}{*}{Classifier} & \multirow{2}{*}{NCh} & \multicolumn{2}{|c|}{10-fold Acc in \%} & \multicolumn{2}{c}{10-fold Sn, Sp in \%} \\ \cline{3-6} 
                     & & En & EnD & En & EnD \\ \hline \hline

                     \multirow{3}{*}{ENS} & 1 & 89.46 & 89.26 & 82.58, 94.08 & 86.08, 91.75 \\
                     & 2 & 93.16 & 96.68 & 89.03, 96.40 & 95.19, 97.84 \\
                     & 3 & 95.58 & 97.68 & 92.93, 97.67 & 96.81, 98.38 \\ \hline

                     \multirow{3}{*}{k-NN} & 1 & 83.63 & 84.84 & 81.28, 85.47 & 80.80, 87.98 \\
                     & 2 & 93.47 & 93.77 & 92.46, 94.27 & 92.23, 94.99 \\
                     & 3 & 96.08 & 96.68 & 95.44, 96.58 & 96.12, 97.12 \\ \hline

                     \multirow{3}{*}{SVM} & 1 & 83.32 & 85.94 & 83.12, 83.50 & 86.03, 85.83 \\
                     & 2 & 90.56 & 92.07 & 92.02, 89.43 & 91.55, 92.49 \\
                     & 3 & 90.25 & 92.16 & 91.78, 89.24 & 92.43, 91.93 \\ \hline

                \end{tabular} }
                
                \label{t-acc5-wav}
            \end{table}

            \begin{table}[h!]
                \centering
                  \caption{Results obtained with SLBP-based features using 10-fold cross-validation strategy.}
                \resizebox{1.1\textwidth}{!}{
                \begin{tabular}{  c | c | C{1.5cm} | C{1.5cm} | C{3.0cm} | C{3.0cm} } 
                     \hline
                     \multirow{2}{*}{Classifier} & \multirow{2}{*}{NCh} & \multicolumn{2}{|c|}{10-fold Acc in \%age} & \multicolumn{2}{c}{10-fold Sn, Sp in \%age} \\ \cline{3-6} 
                     & & En & EnD & En & EnD \\ \hline \hline

                     \multirow{3}{*}{ENS} & 1 & 81.52 & 85.44 & 75.10, 86.55 & 82.19, 87.99 \\
                     & 2 & 87.45 & 94.87 & 83.55, 90.50 & 92.92, 96.40 \\
                     & 3 & 91.67 & 96.48 & 88.83, 93.90 & 95.88, 96.95 \\ \hline

                     \multirow{3}{*}{k-NN} & 1 & 83.54 & 84.43 & 81.05, 8549 & 78.34, 89.25 \\
                     & 2 & 95.38 & 97.89 & 94.30, 96.24 & 96.58, 98.92 \\
                     & 3 & 99.19 & 99.29 & 98.86, 99.64 & 99.09, 98.92 \\ \hline

                     \multirow{3}{*}{SVM} & 1 & 85.54 & 86.44 & 82.64, 87.81 & 82.39, 89.61 \\
                     & 2 & 92.06 & 96.38 & 88.57, 94.79 & 94.30, 98.03 \\
                     & 3 & 95.98 & 97.79 & 95.43, 96.40 & 97.48, 98.02 \\ \hline

                \end{tabular} }
              
                \label{t-acc5-slbp}
            \end{table}

\subsection{Discussion}

Considering the results obtained from all three validation schemes, it may be noted that the wavelet-based method and SLBP consistently outperformed the EMD-based approach. Interestingly, the SLBP-based method produced accuracy results comparable to the wavelet-based technique, while being computationally more efficient. This is because the SLBP method does not require signal decomposition but operates on neighboring samples, and it does not necessitate the computation of additional features as required in the wavelet and EMD methods. Also, it can be inferred that EnD-based channel selection is effective in selecting the channels useful in ADHD detection. 

\begin{figure*}[!h]
            \centering
            \centerline{\includegraphics[scale=.5]{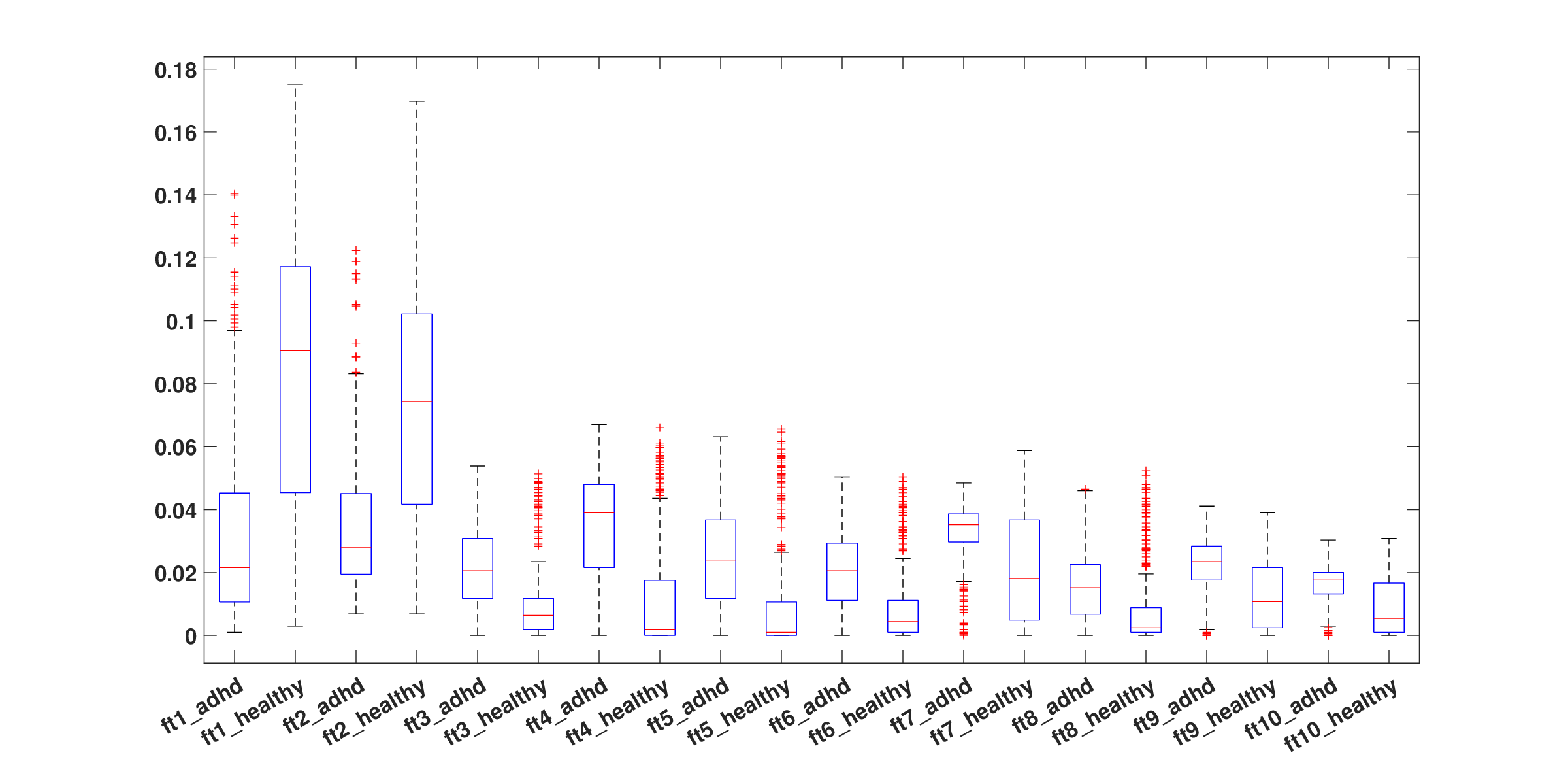}}
            \caption{Box plot of first 10 SLBP-based features obtained from first 3 EnD channels of the training set.}
            \label{figboxtr}
            \end{figure*}

\begin{figure*}[!h]
            \centering
            \centerline{\includegraphics[scale=.5]{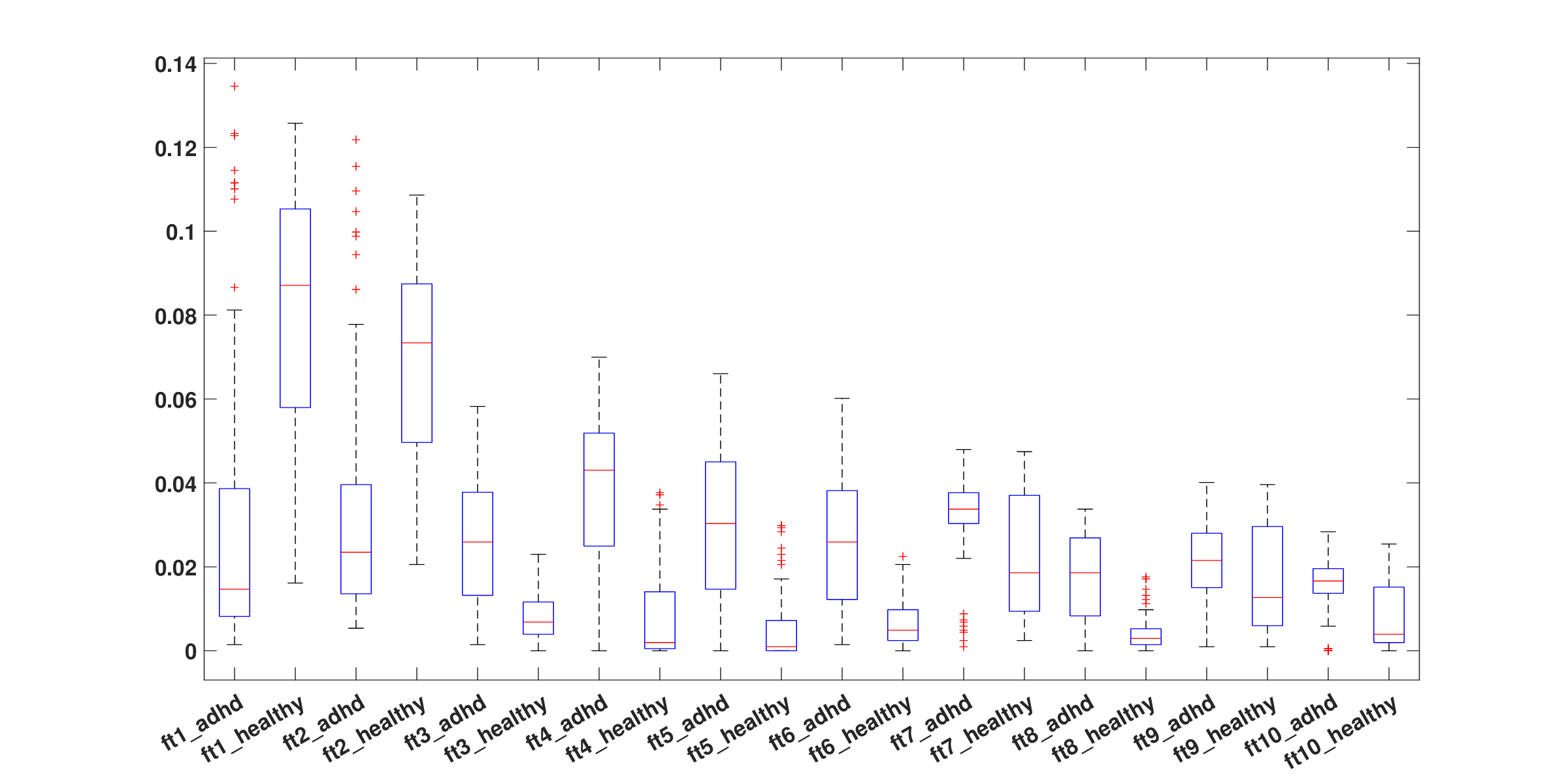}}
            \caption{Box plot of first 10 SLBP-based features obtained from first 3 EnD channels of the testing set.}
            \label{figboxts}
            \end{figure*}

To assess the discriminative ability of features extracted from EnD-based channels, we conducted a statistical analysis (with only SLBP features). Box plots were generated for the first 10 SLBP features, representing the training and testing sets, as depicted in Figures~\ref{figboxtr} and \ref{figboxts}, respectively. These plots clearly demonstrated that the median values of these features are distinct and well separated between the two sets.
            
The performance comparison of the proposed approach with existing EEG-based ADHD detection methods is shown in Table~\ref{t-comp}. The table shows that the proposed approach has outperformed the existing approaches in ADHD detection.  The work in~\cite{r1} tried to mimic a deep learning model by using TMP and tunable Q-wavelet transform (TQWT) for extracting the significant features. However, employing a subband decomposition scheme imposed additional complexity to the model. They obtained the highest accuracy using ten-fold cross-validation. In work~\cite{r2}, the EEG segments were decomposed into 85 subband coefficients using TQWT and wavelet packet decomposition (WPD) techniques. Later, 86 handheld feature set is extracted from the subbands. Again, it is a complex model that involves time-frequency approaches and a vast feature generation process. The authors also reported their highest accuracy using a 10-fold cross-validation scheme. The approach in~\cite{r3} also employed EEG signal segmentation into its individual frequency bands for feature extraction. They also employed ANN and genetic algorithms for ADHD classification. The work in~\cite{r4} employed deep learning methods for classification purposes. Employing deep neural networks has limitations like increased computational complexity and lack of generalization when training data is limited.
The primary advantages, limitations, and future works are discussed below:
\subsubsection*{Advantages}
\begin{itemize}
    \item Proposed approach requires features from only a subset of EEG channel. Hence, it effectively reduces the computational burden compared to using all channels. Enables faster processing and real-time applications, enhancing practical usability.
    \item Despite using a reduced number of channels, the approach maintains the performance of ADHD detection.
    \item The use of EnD-based channel selection offers an objective and systematic method for identifying the most informative EEG channels. Helps in eliminating subjective biases in channel selection.
\end{itemize}
\subsubsection*{Limitation}
\begin{itemize}
    \item The main limitation of our approach is that the number of channels to be used must be manually specified.
\end{itemize}
\subsubsection*{Future Scope} 
\begin{itemize}
    \item The future direction involves incorporating deep learning techniques to further enhance the performance of the approach.
    \item The EnD-based channel selection technique can be employed for autism, epilepsy, Parkinson's disease, etc.
    \item Future efforts could focus on automating the determination of the optimal number of EEG channels required. This can be achieved by developing algorithms that adaptively adjust the channel count based on data characteristics.
    \item We also plan to detect ADHD using electrocardiogram (ECG) signals using our proposed method ~\cite{koh2022automated}.
    \item Also, explainable artificial intelligence (XAI) and uncertainty quantification can be used to evaluate the influence of noise on the developed model ~\cite{jahmunah2023uncertainty}.
\end{itemize}

            \begin{table}[h!]
                \centering
                 \caption{Performance comparison with existing authomated ADHD detection approaches developed.}
                \resizebox{1.1\textwidth}{!}{
                \begin{tabular}{  c | c | c | c } 
                     \hline
                     Paper & Approach & Accuracy (\%) & Dataset used \\ \hline \hline

                    \cite{r1} & Tunable-Q wavelet transform-based & 95.57 & 61 ADHD  \\
                    & features along with kNN classifier &  & 60 HC \\ \hline

                    \cite{r2}& Eight-pointed star pattern  & 97.19 & 61 ADHD  \\
                    & learning network & & 60 HC \\ \hline

                    \cite{r3}& Directed Phase Transfer   & 89.2 & 61 ADHD  \\
                    & Entropy (dPTE) based approach & & 60 HC \\ \hline

                    \cite{r4}& RGB Image extracted from  & 97.47 & 61 ADHD  \\
                    & frequency bands are fed to CNN & & 60 HC \\ \hline

                    \cite{r5}& Non-linear features extracted from  & 96.05 & 50 ADHD  \\
                    & EEG signals are fed to SVM classifier & & 26 HC \\ \hline

                    \cite{r6}& nCREANN & 99 & 61 ADHD  \\
                    &  & & 60 HC \\ \hline


                    - & Proposed approach & 88.52 (70:30 split) & 61 ADHD  \\
                    &  & 95.68 (70:30 split performed 10 times) & 60 HC \\
                    &  & 99.29 (10-fold cross-validation) & \\ \hline

                \end{tabular} }
               
                \label{t-comp}
            \end{table}

\section{Conclusion}
\label{sec.4}
    In this paper, an automated approach has been developed to detect ADHD using an EnD-based channel selection algorithm. Our EnD-channel selection-based approach achieved an accuracy of 99.19\% for the automated identification of ADHD. More specifically, classification accuracy is obtained when only the first three significant channels are used for classification purposes. This channel selection approach significantly reduces the computational complexity. Also, the channels selected using the EnD-based channel selection approach were found to be more effective than the channels selected by entropy-based channel selection in the automated ADHD detection.  

    In this paper, we have presented EnD-based channel selection for ADHD detection. The EEG channel selection is one of the problems in the EEG classification. Therefore, in future, we plan to explore the proposed approach for the identification of significant channels for schizophrenia detection, epilepsy detection, and brain-computer interface using huge EEG database obtained from many centers. 
\section*{Conflicts of interest/Competing interest}
The authors have no relevant financial or non-financial interests to disclose.

The authors have no conflicts of interest to declare that are relevant to the content of this article.

All authors certify that they have no affiliations with or involvement in any organization or entity with any financial interest or non-financial interest in the subject matter or materials discussed in this manuscript.

The authors have no financial or proprietary interests in any material discussed in this article.


\bibliographystyle{elsarticle-num} 
\bibliography{paper_tex}

\end{document}